\documentclass[a4paper,12pt,english,german]{article}

\usepackage{mathtools}
\usepackage{graphicx}

\begin{document}

\centerline{\bf  Bound States as Emergent Quantum Structures.}
\bigskip

\centerline{ R. E. Kastner\footnote{Foundations of Physics Group, University of Maryland, College
Park, USA;  rkastner@umd.edu}}
 \centerline{26 January 2016}

\bigskip

{\small ABSTRACT. Bound states arise in many interactions among elementary field states, and are represented by poles in the scattering matrix. The emergent nature of bound states suggests that they play a perhaps under-appreciated role in specifying the ontologically relevant degrees of freedom pertaining to composite systems. The basics of this ontology are presented, and it is discussed in light of an example of Arsenijevi\' c et al.}

\bigskip

\noindent {\bf 1. Bound states: a definition}\\

Suppose we are given two or more interacting degrees of freedom described by some field Lagrangian. One can apply scattering theory to these degrees of freedom, and for attractive interactions certain final states will be bound states, which establish a new composite entity.

 The paradigmatic example is the hydrogen atom. In the nonrelativistic description, one can treat the proton as the generator of a potential well and the electron as an incoming wave function. A scattering matrix can be defined, which assigns amplitudes to various outgoing states for the particles. For suitable incoming momenta of the electron, some of those states will correspond to the situation in which the scattered electron state has imaginary energies; these will result in an exponentially decaying wave function that traps the electron in the potential well due to the proton. These are the bound states, which result in the creation of the hydrogen atom as a new composite entity. The bound states (and resonances which are metastable bound states) correspond to poles in the scattering matrix for the interacting quanta.\footnote{See \cite{Weinberg}, p. 431 for a careful discussion of this.} Thus, for purposes of this discussion we define the bound state as a final scattering state characterized by a pole in the scattering matrix describing an encounter between two or more degrees of freedom.\footnote{Technically we need to add that the potential modeling the interaction falls off faster than any exponential as separation of the degrees of freedom approaches infinity.}
 \bigskip

\noindent {\bf 2. Bound states as emitters and absorbers}\\

A bound system such as a hydrogen atom is of course subject to many different energy levels, corresponding to the different possible bound states for the electron. An excited state is subject to decay to a lower state, upon which the electron emits a photon. In the transactional interpretation (TI) of Cramer\cite{Cramer}, at least one additional absorbing bound system is required as a condition for a real photon to be emitted from an excited atom. This is because TI is based on a direct-action theory, in which radiation occurs as a result of a mutual direct interaction between emitter and the absorber. The classical version of the direct-action theory was developed by Wheeler and Feynman \cite{WF} and the quantum relativistic version by Davies\cite{Davies}. 

TI defines the usual quantum state $| \Psi \rangle$ as an `offer wave' (OW), and it defines the advanced response $\langle a|$ of an absorber $A$ as  a `confirmation wave' (CW). In general, many absorbers $A,B,C,....$ respond to an OW, where each absorber responds to the component of the OW that reaches it. The OW component reaching an absorber $X$ would be $\langle x | \Psi \rangle |x\rangle$, and it would respond with the adjoint (advanced) form $\langle x | \langle \Psi |x\rangle$. The product of these two amplitudes corresponds to the final amplitude of the `echo' of the CW from $X$ at the locus of the emitter and corresponds to the Born Rule. Meanwhile, the sum of the weighted outer products (projection operators) based on all CW responses corresponds to the mixed state identified by von Neumann as resulting from the non-unitary process of measurement.  Thus, TI provides a physical explanation for both the Born Rule and the measurement transition from a pure to a mixed state.

Due to energy conservation, a free charged particle can neither emit nor absorb a photon; so bound states are the true emitters  and absorbers. The present author has proposed a relativistic extension of TI, called the possibilist transactional interpretation (PTI) \cite{Kastner}. PTI regards bound states as important and fundamental quantum structures. According to PTI, spacetime is a discrete manifold that is emergent from a quantum level of physical possibilities, which are the quantum degrees of freedom (including bound states). It is transactions between an emitting bound state and an absorbing bound state that generate spacetime events and their connections: the emitter and absorber define the endpoints, and the transferred quantum of conserved quantities (energy, momentum, spin, etc.) defines their connection. Thus, when an electron in an atom emits a photon, the entire atom is actualized as a spacetime object, and similarly, the entire absorbing atom is actualized as a spacetime object as well. This is because, in each case, the entire atom undergoes a well-defined physical change of state as a result of the transaction; it is not just the component electron that is affected.
\bigskip

\noindent {\bf 3. A physical criterion for structurally significant degrees of freedom}\\

In this volume, Arsenijevi\' c et al \cite{ADD} discuss the challenge of deciding on physically relevant decomposition of composite systems into subsystems. Analyzing the example of an electrically neutral atom in a Stern-Gerlach apparatus, they point out that if the action of the field on the atom is indeed correctly modeled by the usual Hamiltonian interaction term inducing entanglement between the atom's center of mass and its spin, the external magnetic field of the S-G acts on the center of mass of the atom as a whole, and not individually on the electrons and nucleus. For if the latter were the case, the internal state of the atom would be observably changed after passing through the field. 

According to PTI, this is because the atomic bound state is an ontological whole, an emergent structure. Its status as a new, nonseparable entity corresponds to the fact that it is described by a pole in the scattering matrix for its constituent subsystems. Such a singularity in the theoretical description is often an indicator of an emergent process, in which one model breaks down and must be superseded by another.  For example, in quantum field theory a real quantum can be seen as an emergent structure as follows:  the propagator function, defined over the complex frequency plane, describes virtual quantum propagation. A pole in the propagator describes a real quantum, which differs physically from a virtual quantum since the former is representable by a Fock state (stable field excitation) but the latter is not. Similarly, the bound state is not simply an entangled state of two or more degrees of freedom, in which the entanglement can be viewed as relative to a choice of observable. Rather, it is an emergent structure arising from a particular kind of interaction among its constituents, in which they lose their separate identity as free field excitations and form a new collective final state.  

In this picture, clearly the dynamical conserved quantities are considered the fundamental ``preferred'' observables, while the position observable (characterizing an aspect of spacetime phenomena) is secondary and emergent. One might ask why there should be such an asymmetry between the observables. This asymmetry is naturally found at the relativistic level, since at that level there is no well-defined position observable (and time is not even an observable at the non-relativistic level). In contrast, energy, momentum, spin, etc are all well-defined observables at the relativistic level. And it is at the relativistic level where one finds emission (creation of field quanta) and absorption (destruction of field quanta) which must always occur in any measurement context, i.e., whenever it is possible to unambiguously apply a determinate value of an observable to a system. Thus, the dynamical conserved quantities merit consideration as the naturally preferred observables, and this in turn allows for a fundamental structural criterion. 

Finally, it should be noted that one can still obtain some information about the components of bound states. One can image to some extent the component degrees of freedom through probing the shape of the scattering potential they create. The composite system is used as a target, and a probe system is scattered from it.\footnote{Here we consider only elastic collisions, since inelastic ones alter or destroy the internal structure such that a collective degree of freedom would not accurately describe it anyway.} But there is no quantum entanglement between the probe system and the target components: the probe does not 'see them'  individually at the level of their quantum states, but only collectively through the overall shape of the target potential.  An early example is Rutherford's gold foil experiment, in which alpha particle were used as probes for the gold atoms.
\bigskip

\noindent {\bf 4. Conclusion}\\

It has been argued that bound states, as represented by poles in the scattering matrix for their constituents, constitute emergent non-separable structures that are properly characterized by degrees of freedom reflecting their wholeness, such as the center of mass degree of freedom used to model atoms interacting with a Stern-Gerlach field. The entanglement of the composite degrees of freedom of a bound state is of a fundamentally different nature than than of non-bound degrees of freedom, in that it is not relative to a choice of observable, but is an intrinsic, formative property of the entire structure. The bound state's description in terms of a center-of-mass degree of freedom can thus be viewed as physically unambiguous and descriptive of the composite system in a way that the component degrees of freedom are not. This is yet another demonstration of the signature of the quantum world that the ``whole is greater than the sum of the parts.'' It is also noted that the dynamical conserved quantities are the natural ``preferred observables'' when the relativistic level is taken into account in investigating the origins of quantum structures.
\bigskip

{\bf Acknowledgment}\
I would like to thank Miroljub Dugi\' c and Jasmina Jekni\' c-Dugi\' c for valuable discussions.

\end{document}